# Metasurface Optics for Full-color Computational Imaging


**Authors:** Shane Colburn[1], Alan Zhan[2], Arka Majumdar[1,2]*

**Affiliations:**

[1]Department of Electrical Engineering, University of Washington, Seattle, Washington 98195, USA.

[2]Department of Physics, University of Washington, Seattle, Washington 98195, USA.

*Correspondence to: arka@uw.edu



**Abstract**: Conventional imaging systems comprise large and expensive optical components which successively mitigate aberrations. Metasurface optics offers a route to miniaturize imaging systems by replacing bulky components with flat and compact implementations. The diffractive nature of these devices, however, induces severe chromatic aberrations and current multi-wavelength and narrowband achromatic metasurfaces cannot support full visible spectum imaging (400-700 nm). We combine principles of both computational imaging and metasurface optics to build a system with a single metalens of NA ~ 0.45 which generates in-focus images under white light illumination. Our metalens exhibits a spectrally invariant point spread function which enables computational reconstruction of captured images with a single digital filter. This work connects computational imaging and metasurface optics and demonstrates the capabilities of combining these disciplines by simultaneously reducing aberrations and downsizing imaging systems with simpler optics.


**One Sentence Summary:** We design metalenses to capture full visible spectrum color images with white-light by combining metasurface optics and computational imaging.



**Main Text:**

**Introduction**

Modern cameras consist of systems of cascaded and bulky glass optics for imaging with minimal aberrations. While these systems provide high quality images, the improved functionality comes at the cost of increased size and weight, limiting their use for a variety of applications requiring compact image sensors. One route to reduce a system's complexity is via computational imaging, in which much of the aberration correction and functionality of the optical hardware is shifted to post-processing in the software realm, enabling high quality images with significantly simpler optics (*1, 2*). Alternatively, the designer could miniaturize the optics by replacing them with diffractive optical elements (DOEs), which mimic the functionality of refractive systems in a more compact form factor. Metasurfaces are an extreme example of such DOEs, in which quasiperiodic arrays of resonant subwavelength optical antennas impart spatially-varying changes on a wavefront (*3–5*). These elements are of wavelength-scale thickness, enabling highly compact systems, while the large number of degrees of freedom in designing the subwavelength resonators has enabled unprecedented functionalities and flat implementations of lenses (*6–12*), holographic plates (*13, 14*), blazed gratings (*15, 16*), and polarization optics (*17*).

While separately both computational imaging and metasurfaces are promising avenues toward simplifying optical systems, a synergistic combination of these fields can further enhance system performance and facilitate advanced capabilities, for example, for use in full visible spectrum imaging with metasurfaces. Designing achromatic metasurface lenses for imaging under broadband illumination remains an outstanding problem in the metasurface community. The strong chromatic aberrations in metasurfaces originate from both the local resonant behavior of the sub-wavelength optical scatterers, as well as from phase wrapping discontinuities arising



from the spatial arrangement of the scatterers (*18*). For lenses, this chromaticity manifests as wavelength-dependent blur in images, which constrains metasurface-based imaging to narrowband operation (*19–21*). There is a vast body of work attempting to solve this problem; however, thus far the presented solutions either work for discrete wavelengths (*18, 22–25*) or narrow bandwidths (*26, 27*). Here, we design the optical hardware in conjunction with computational post-processing to realize a full-color imaging system comprising a single metasurface and a computationally inexpensive digital filter which can generate high quality images under broadband white light illumination spanning the whole visible regime.

**Results**

The 3D point spread function (PSF) of a linear, shift-invariant optical system fully characterizes its behavior. At the image plane of an optical system, the 2D PSF corresponds to an image of a point source with size and shape related to the system's geometry and aberrations. As the wavelength changes, the image plane shifts due to chromatic aberrations, inducing color-dependent blur in captured images because of the fixed location of the image sensors. For metasurface optics, this focal shift is inversely proportional to the optical wavelength, severely blurring polychromatic images. We mitigate this blur by engineering a metasurface with a PSF which is invariant across the whole visible regime. This technique has been explored previously with macroscopic refractive optical systems, in which the optical wavefront is coded using a phase mask to provide an extended depth of focus (EDOF) (*28–30*). This EDOF makes the system tolerant to focal shifts due to the preservation of spatial frequency information across the depth of the smeared out focal spot. This comes with the tradeoff of reduced signal-to-noise ratio (SNR) and blurring of the captured image as incident light is spread over a greater volume; however, unlike a simple lens which has wavelength-dependent blur, an EDOF system can have



spectrally invariant blur over a wide frequency band, the bandwidth of which increases with the depth of focus (see supplementary materials for a performance comparison of metasurface systems with different depths of focus). The spectral invariance of the induced blur enables post-processing with a single wavelength-independent filter to retrieve a high-quality image.

The requirement of a secondary phase mask for wavefront coding increases the system's size and complexity. Furthermore, these phase masks are often freeform in nature (i.e., characterized by rotational asymmetry or higher order polynomials) and are challenging to fabricate by traditional means such as diamond turning and multi-stage lithography for making diffractive elements. With the flat nature of metasurface-based systems, however, we can convert a freeform element to a compact and uniform thickness device using a single lithography stage (*31*). This design freedom also enables combining the lensing and wavefront coding functionalities into a single element. A variety of different wavefront coding masks can produce an EDOF. Typically, these masks will produce non-diffracting beams over a wide range and in conjunction with a lens can generate an elongated focus. Cubic functions are a common choice because of their simplicity and rectangular separability (*30*). Furthermore, it can be shown via the stationary phase method applied to an ambiguity function representing an optical transfer function, that for a phase mask which is a monomial, the modulation transfer function is insensitive to misfocus (e.g., chromatic focal shift) if and only if the mask function is cubic (*29*). Owing to this wavelength insensitivity and their widespread usage in extended depth of field systems, we have elected to use a cubic phase term in this work as well. Here, we design a single element capable of simultaneously focusing light and coding the wavefront to increase the depth of focus with phase of the form below:

$$\varphi = \frac{2\pi}{\lambda} \left( \sqrt{x^2 + y^2 + f^2} - f \right) + \frac{\alpha}{L^3} (x^3 + y^3)$$



where x and y are the in-plane coordinates, λ the operating wavelength, f the nominal focal

length, L half the aperture width, and α signifies the extent of the cubic phase. We designed two

metasurfaces: one with α = 0 which is a simple lens, and one with α = 55π, which has an EDOF.

Both devices had a nominal focal length of 200 μm at 550 nm. Our devices consist of cylindrical

silicon nitride nanoposts of thickness 633 nm on top of a silicon dioxide substrate (Figure 1A-B)

positioned on a square lattice with a period of 400 nm. Our choice of nanoposts, as opposed to

nanofins (*10*, *19*) or V-shaped antennas (*5*), enables polarization-independent behavior, while the

high bandgap of silicon nitride enables transparent operation and high efficiency across the

visible band (*32*). Each nanopost mimics a truncated waveguide with low reflectivity top and

bottom interfaces which induce low quality factor resonances. Incident light couples into modes

supported by the nanoposts, which then shift the phase of the light before coupling the light to a

transmitted free-space mode. By adjusting the diameter, the modal structure supported by the

nanopost can vary, changing the modes' ensemble behavior and inducing different phase shifts.

Due to our interest in broadband visible regime operation (400-700 nm), we selected a central

nominal design wavelength of 550 nm for our nanoposts. We simulated the transmission

coefficient as a function of post diameter via rigorous coupled-wave analysis (RCWA) (*33*)

(Figure 1C). The transmission coefficient exhibits more than a $2\pi$ change in phase and uniform

amplitude over a wide range. In designing our metasurfaces, we used a set of 10 different phase

steps between 0 to $2\pi$ and avoided drops in amplitude by selecting diameters off the resonance

dips. We simulated both metasurfaces to analyze their performance as a function of wavelength.

Figures 1D and 1E show the chromatic focal shift for the α = 0 and α = 55π designs respectively,

in which the dashed black lines indicate the desired focal plane. We see that while only green

light (550 nm) is in focus for α = 0, all simulated wavelengths impinge on the desired plane as



part of an EDOF for $\alpha = 55\pi$. We fabricated the metasurfaces using electron-beam lithography and dry etching. Figures 1F-G show optical images of the final devices (see supplementary materials for scanning electron micrographs of the nanoposts). The fabricated $\alpha = 0$ and $\alpha = 55\pi$ lenses demonstrate average measured focusing efficiencies of 63% and 57% respectively (see supplementary materials) for the wavelengths tested, comparable to existing visible wavelength metalenses. Hereafter, we denote the $\alpha = 0$ metasurface as the singlet metalens and the $\alpha = 55\pi$ metasurface as the EDOF metalens.



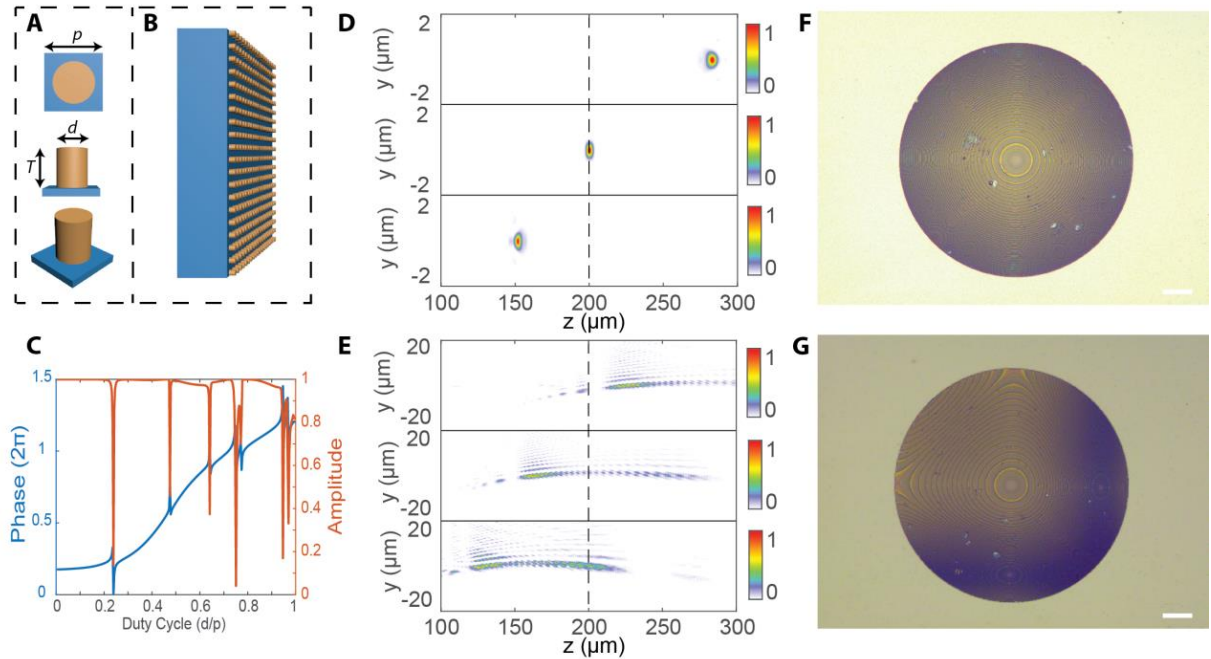

Fig. 1 Design, simulation, and fabrication of imaging metasurfaces. (A) The metasurfaces are made up of silicon nitride nanoposts, where the thickness T, lattice constant p, and diameter d are the design parameters. (B) Schematic of a metasurface comprising an array of nanoposts. (C) Simulation of the nanoposts' transmission amplitude and phase via rigorous coupled-wave analysis. Simulated intensity along the optical axis of the singlet metasurface lens (D) and extended depth of focus metasurface (E) where going from top to bottom in each panel 400, 550, and 700 nm wavelengths are used. The dashed lines indicate the desired focal plane, where the sensor will be placed. Optical images of the singlet metasurface lens (F) and the extended depth of focus device (G). Scale bars are 25 μm.



Our computational imaging system poses a problem of the matrix form f = Kx + n (*34*), where the desired image x has been blurred by the system kernel K and corrupted by noise n to produce the captured image f. A variety of different methods are available for estimating x, such as the linear Wiener filter or regularized optimization-based approaches. For this work, we chose the Wiener filter due to its low computational complexity. Moreover, the reconstructed image quality is comparable to that of more advanced deconvolution methods (see supplementary materials for performance comparisons with total variation-regularized deconvolution). We obtain the kernel K, as required for filtering the image, by a calibration PSF measurement. We measured the PSFs and calculated the modulation transfer functions (MTFs) for both the singlet and EDOF metalenses for red (625 nm), green (530 nm), and blue (455 nm) LED illumination (Figure 2) (see the supplementary materials for further information on the MTF measurement and comparison with theory and illumination with different bandwidth sources). For the singlet, the tightly focused spot with green illumination differs drastically from the large blurs under red and blue illumination, translating to the strong distinction in the MTFs under these illumination conditions. With this design, the zeros in the MTFs for red and blue wavelengths result in an unrecoverable loss of spatial frequency information and preclude computational reconstruction. For the EDOF metasurface, however, not only are the MTFs wavelength-invariant, but they also have no zeros until the cutoff spatial frequency. The preservation of this spatial frequency content enables computational reconstruction to retrieve the desired image x.



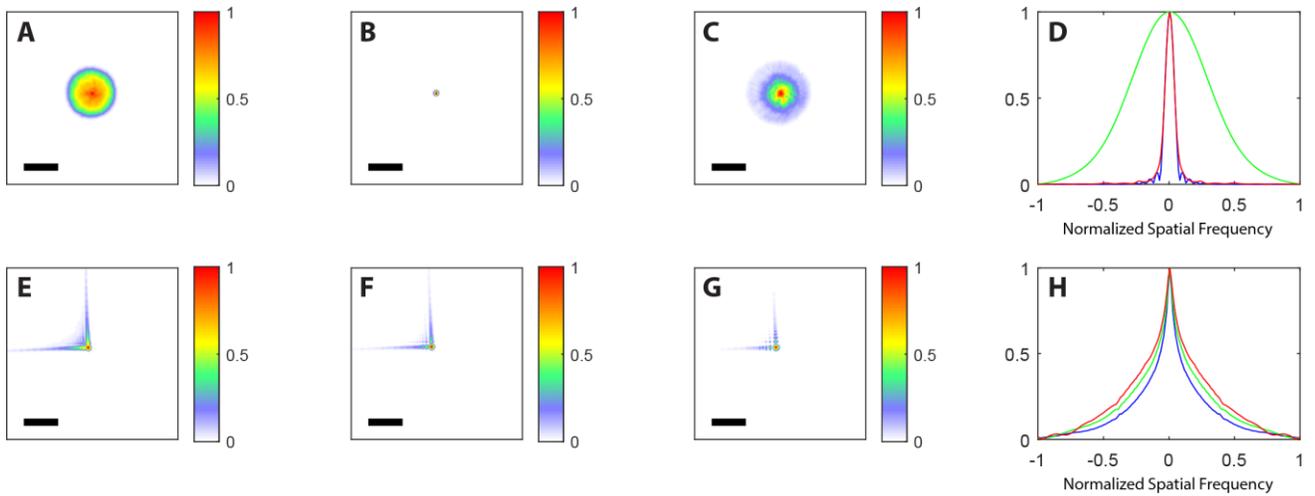

Fig. 2 Characterization of the imaging metasurfaces. The PSFs of the singlet metalens (top row) and extended depth of focus lens (bottom row) were measured under blue (A and E), green (B and F), and red (C and G) illumination conditions. The scale bars are of length 25 μm. The MTFs were also calculated for both designs (D and H). In both D and H, a normalized spatial frequency of 1 corresponds to the same cutoff frequency of 579 cycles/mm.



To demonstrate the imaging capability of the system, we illuminated patterns on standard printer paper at object distances much greater than the focal length, on the order of a few centimeters. The metasurfaces would then form images by focusing down the scattered light. We first examined narrowband (~30nm bandwidth) imaging performance, under separate red (625 nm), green (530 nm), and blue (455 nm) illumination with LEDs. Figures 3C-E compare the captured images of a portion of the 1951 Air Force resolution chart (Figure 3A) by the singlet lens to those by the EDOF lens with and without deconvolution. While the pattern is in focus for green light, upon switching to blue or red illumination, the images captured via the singlet undergo severe distortion. For the EDOF lens without deconvolution, the image is blurry at all wavelengths, but the blur appears uniform across wavelengths. With post-capture deconvolution, however, the resulting images appear in focus for all wavelengths, providing a substantial performance improvement with greatly reduced chromatic aberrations relative to those of the singlet. We quantified this performance improvement in terms of the structural similarity (SSIM) (*35*) of the Air Force pattern images. The deconvolved EDOF system's images had a SSIM 0.209 higher compared to those of the $\alpha = 0$ system (see supplementary materials for further details of this calculation). We also imaged more complex patterns, including a black and white binary Mona Lisa pattern (Figure 3B and 3F-H) and also observed mitigation of the chromatic aberrations using the EDOF lens. While the SNR of our deconvolved images is lower compared to that of the in-focus green images with the singlet metalens, our system exhibits in-focus images over a broad wavelength range. In conjunction with computational algorithms optimized for low-light imaging scenarios comparable to our experimental setup, imaging performance can be further improved.



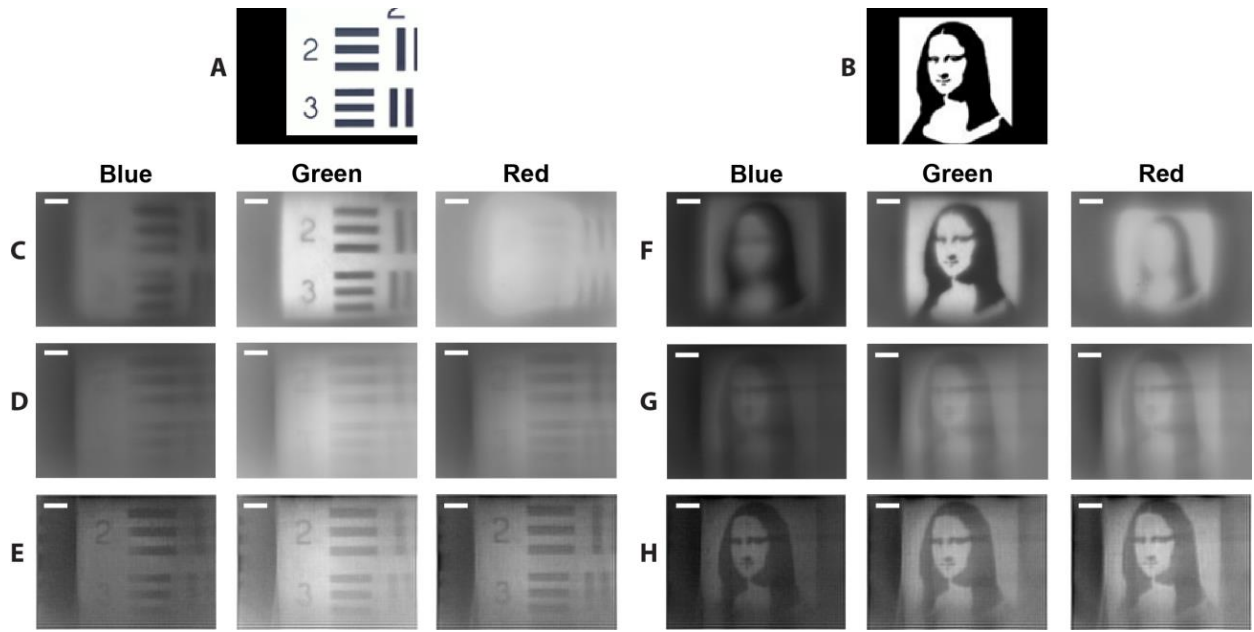

Fig. 3 Imaging at discrete wavelengths. The appropriately cropped original object patterns used for imaging are shown in (A) and (B). Images were captured of the 1951 Air Force resolution chart with the singlet metalens (C) and the EDOF lens without (D) and with deconvolution (E). Images were also taken of a binary Mona Lisa pattern with the singlet metalens (F) and the EDOF device without (G) and with (H) deconvolution. The scale bars are of length 20 μm.



Finally, we tested the system under broadband illumination using a white light source (Figure 4). Under this condition, the singlet lens significantly blurred color printed RGB text (Figure 4A), with noticeably better image quality for the green letter G as the lens was positioned to be in focus for green light. Figure 4A also shows the spectrally uniform blurring of RGB text formed by capturing the same object pattern with the EDOF lens. After deconvolution, we can clearly make out each individual character, whereas the blue B is blurry and the red R is unintelligible in the image captured directly with the singlet lens. The deconvolved images do exhibit some erroneous horizontal and vertical lines arising from the asymmetric shape of the PSF, which can produce directional artifacts, but these can be corrected by more advanced deconvolution as well as by using a rotationally symmetric PSF (*36*). Figure 4B demonstrates a similar image quality improvement for ROYGBIV text, where the characters are substantially blurred by the singlet metalens, but are in focus after capturing with the EDOF device and deconvolution. For the case of a rainbow pattern (Figure 4C), the chromatic blur induced by the singlet obscures individual color bands and the green stripe is barely evident, whereas the deconvolved EDOF image clearly shows separate bands and edges. For a landscape image (Figure 4D) with multicolor flowers and leaves, stem and leaf structures severely blurred by the singlet are in focus and color ringing artifacts in the blossoms are reduced in the deconvolved EDOF image.



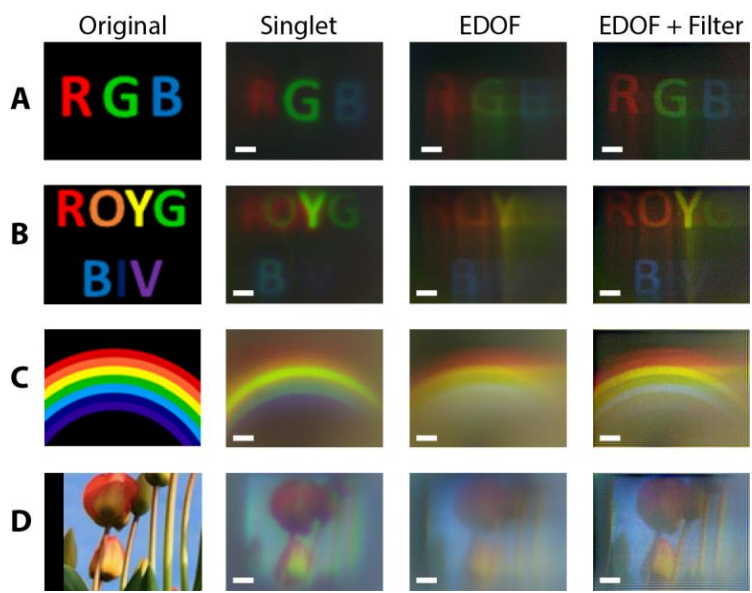

Fig. 4 Imaging with white light. Images were taken under incoherent white light illumination of color printed RGB (A) and ROYGBIV (B) text, a colored rainbow pattern (C), and picture of a landscape (D) with a blue sky, green leaves, and multicolor flowers, using a color camera. The appropriately cropped original object patterns used for imaging are shown in the left column. The scale bars are of length 20 μm.



**Discussion**

Compared to metasurface designs where the achromatic bandwidth is narrow (*26, 27*) or discrete wavelength images are superposed to produce a color image (*24*), to the best of our knowledge, we are the first to demonstrate in-focus full visible spectrum imaging with white light directly. We observe this behavior in Fig. 4, where when a single image is captured we have not only red, green, and blue light in focus after deconvolution, but also intermediate colors, such as yellow, orange, and violet. Furthermore, our metalenses rely on simple transmissive scatterers which could be extended to use a variety of different geometries and materials, whereas existing broadband achromatic metalenses require scatterers with carefully designed dispersion characteristics (*26, 27*). With our system, however, while the digital filter in conjunction with our modified phase mask enables broadband color imaging, the required post-processing complicates the system and introduces a delay time to deconvolve the captured image. For many photography and video applications this would not pose an issue as any captured frames could be saved and then deconvolved offline. For real-time imaging, our system would also work as our filter relies on the $O(NlogN)$ fast Fourier transform algorithm, which can be accelerated via FPGAs or GPUs. Such hardware acceleration techniques would require additional circuitry and increase system complexity, requiring the designer to balance the system's requirements and costs for the given application. Our implemented system also suffers from a limited space-bandwidth product arising from our small (200 μm) focal length and aperture width. Typically, the space-bandwidth product of an optical system will decrease as the system dimensions are scaled down, reducing the information capacity and number of resolvable points in an image (*37*). The small aperture of our metalenses also limits light collection, which reduces the SNR and necessitates higher incident power or increased exposure time. These limitations, however, are not inherent to our



hybrid optical-digital system, but instead they arise from our short focal length and would be present in any implementation at the same length scale.

The reported system combines computational imaging and an EDOF metasurface with a NA ~ 0.45 to image full visible spectrum object patterns with minimal chromatic aberrations, making our devices well-suited for microscopy, hyperspectral imaging, and ultra-thin cameras. To the best of our knowledge, we have also imaged with the shortest focal length metasurface to date with more than a 2.4X reduction in focal length compared to the shortest value previously (*26*). While this system must contend with geometric aberrations, these can be circumvented by further co-optimization of the optical element (e.g., by stacking metasurfaces (*21, 38*)) and post-processing algorithms (*39*). In combining optical metasurfaces and computational imaging, this work provides a model for designing hybrid systems where the optical hardware and software together generate high quality images while minimizing system size and complexity.

**Material and Methods**

<u>Design and simulation</u>

The nanoposts were designed to give 0 to $2\pi$ phase for wavelengths spanning the entire visible regime (400-700 nm) as the diameter of the posts were varied. Figure S1 shows the transmission amplitude and phase for three representative wavelengths at 400 nm, 550 nm, and 700 nm, calculated via rigorous coupled-wave analysis (RCWA) simulation. The refractive indices used for both the silicon nitride posts and the silicon dioxide substrate accounted for dispersion. To ensure our nanopost design provided weakly coupled pillars that would allow us to make use of the unit cell approximation in implementing our phase profiles, we simulated the transmission amplitude (Fig. S2A) and phase (Fig. S2B) as a function of diameter as we swept the lattice constant to show minimal change in phase over a wide range of lattice constants. To



design the metasurfaces nominally for 550 nm, the RCWA data served as a lookup table for mapping the phase at a given position to the post diameter which most accurately provides the desired phase. Due to the large spatial extent of our designs, the memory requirements were beyond our available computational resources to perform a full finite-difference time-domain (FDTD) simulation of our devices. As such, we used the RCWA-calculated transmission coefficients and modelled our metasurfaces as complex amplitude masks and simulated their performance by evaluating the Rayleigh-Sommerfeld diffraction integral using an angular spectrum propagator.

<u>Fabrication</u>

A 633 nm film of silicon nitride was first deposited on a fused silica wafer substrate via plasma-enhanced chemical vapor deposition. The wafer was temporarily coated with a protective photoresist layer and diced into smaller pieces before cleaning by sonication in acetone and isopropyl alcohol. The sample was then spin coated with ZEP-520A before sputtering 8 nm of Au/Pd as a charge dissipation layer. The sample was exposed using a JEOL JBX6300FS electron-beam lithography system and the charge dissipation layer was removed by type TFA gold etchant. After developing in amyl acetate, a layer of aluminum was evaporated onto the sample and after performing lift-off, an aluminum hard mask was left on the silicon nitride layer for subsequent etching. The sample was etched using an inductively coupled plasma etcher with a CHF3 and O2 chemistry and the remaining aluminum was removed by immersing in AD-10 photoresist developer. Scanning electron micrographs of the fabricated devices are presented in Fig. S3.



<u>Device characterization</u>

The focal planes of the fabricated metasurfaces were characterized via the experimental setup presented in Fig. S4. Light from a fiber-coupled LED illuminates the metasurface under test and a custom microscope assembled from a translatable stage, objective, tube lens, and camera takes snapshots of the focal plane of the device. To measure the efficiency (Fig. S10), the same setup was used with the addition of a flip mirror, pinhole, and photodetector (Newport 818-SL). The efficiency was calculated by taking the ratio of the power at the focal plane to that of the incident beam (21). The incident beam power was found by measuring the power through a piece of glass with the pinhole aperture set to image a region equal to the width of the metalens. The camera was corrected for dark noise by taking a sequence of calibration images with the lens cap on. We determined the modulation transfer functions (MTFs) of our lenses by Fourier transforming and then taking the magnitude of the measured focal spot.

<u>Imaging</u>

Images were captured using the setup shown in Fig. S5. Light from a fiber-coupled LED is incident off-axis on a pattern printed on standard 8.5 x 11 paper. The metasurface creates an image near its focal plane by focusing light scattered off the printed pattern and a translatable microscope consisting of an objective, tube lens, and camera capture this image. Prior to image capture, the camera's dark noise is subtracted after taking a sequence of pictures with the lens cap on.



# References and Notes:


1.  F. Heide *et al.*, *ACM Trans Graph*, in press, doi:10.1145/2516971.2516974.

2.  C. J. Schuler, M. Hirsch, S. Harmeling, B. Schölkopf, in *2011 International Conference on Computer Vision* (2011), pp. 659–666.

3.  N. Yu, F. Capasso, Flat optics with designer metasurfaces. *Nat. Mater.* **13**, 139–150 (2014).

4.  Planar Photonics with Metasurfaces | Science, (available at http://science.sciencemag.org/content/339/6125/1232009).

5.  N. Yu *et al.*, Light Propagation with Phase Discontinuities: Generalized Laws of Reflection and Refraction. *Science*. **334**, 333–337 (2011).

6.  A. Arbabi, R. M. Briggs, Y. Horie, M. Bagheri, A. Faraon, Efficient dielectric metasurface collimating lenses for mid-infrared quantum cascade lasers. *Opt. Express*. **23**, 33310–33317 (2015).

7.  P. R. West *et al.*, All-dielectric subwavelength metasurface focusing lens. *Opt. Express*. **22**, 26212–26221 (2014).

8.  F. Lu, F. G. Sedgwick, V. Karagodsky, C. Chase, C. J. Chang-Hasnain, Planar high-numerical-aperture low-loss focusing reflectors and lenses using subwavelength high contrast gratings. *Opt. Express*. **18**, 12606–12614 (2010).

9.  A. Arbabi, Y. Horie, A. J. Ball, M. Bagheri, A. Faraon, Subwavelength-thick lenses with high numerical apertures and large efficiency based on high-contrast transmitarrays. *Nat. Commun.* **6**, ncomms8069 (2015).

10. D. Lin, P. Fan, E. Hasman, M. L. Brongersma, Dielectric gradient metasurface optical elements. *Science*. **345**, 298–302 (2014).

11. D. Fattal, J. Li, Z. Peng, M. Fiorentino, R. G. Beausoleil, Flat dielectric grating reflectors with focusing abilities. *Nat. Photonics*. **4**, 466–470 (2010).

12. Aberration-Free Ultrathin Flat Lenses and Axicons at Telecom Wavelengths Based on Plasmonic Metasurfaces - Nano Letters (ACS Publications), (available at http://pubs.acs.org/doi/abs/10.1021/nl302516v).

13. X. Ni, A. V. Kildishev, V. M. Shalaev, Metasurface holograms for visible light. *Nat. Commun. Lond.* **4**, 2807 (2013).

14. G. Zheng *et al.*, Metasurface holograms reaching 80% efficiency. *Nat. Nanotechnol.* **10**, 308–312 (2015).

15. P. Lalanne, S. Astilean, P. Chavel, E. Cambril, H. Launois, Blazed binary subwavelength gratings with efficiencies larger than those of conventional échelette gratings. *Opt. Lett.* **23**, 1081–1083 (1998).

16. S. Astilean, P. Lalanne, P. Chavel, E. Cambril, H. Launois, High-efficiency subwavelength diffractive element patterned in a high-refractive-index material for 633??nm. *Opt. Lett.* **23**, 552–554 (1998).

17. A. Arbabi, Y. Horie, M. Bagheri, A. Faraon, Dielectric metasurfaces for complete control of phase and polarization with subwavelength spatial resolution and high transmission. *Nat. Nanotechnol.* **10**, 937–943 (2015).

18. E. Arbabi, A. Arbabi, S. M. Kamali, Y. Horie, A. Faraon, Multiwavelength polarization-insensitive lenses based on dielectric metasurfaces with meta-molecules. *Optica*. **3**, 628–633 (2016).





19. M. Khorasaninejad *et al.*, Metalenses at visible wavelengths: Diffraction-limited focusing and subwavelength resolution imaging. *Science*. **352**, 1190–1194 (2016).

20. Multispectral Chiral Imaging with a Metalens - Nano Letters (ACS Publications), (available at http://pubs.acs.org/doi/abs/10.1021/acs.nanolett.6b01897).

21. A. Arbabi *et al.*, Miniature optical planar camera based on a wide-angle metasurface doublet corrected for monochromatic aberrations. *Nat. Commun.* **7**, ncomms13682 (2016).

22. Multiwavelength achromatic metasurfaces by dispersive phase compensation | Science, (available at http://science.sciencemag.org/content/347/6228/1342).

23. E. Arbabi, A. Arbabi, S. M. Kamali, Y. Horie, A. Faraon, Multiwavelength metasurfaces through spatial multiplexing. *Sci. Rep.* **6**, srep32803 (2016).

24. O. Avayu, E. Almeida, Y. Prior, T. Ellenbogen, Composite functional metasurfaces for multispectral achromatic optics. *Nat. Commun.* **8**, ncomms14992 (2017).

25. B. Wang *et al.*, Visible-Frequency Dielectric Metasurfaces for Multiwavelength Achromatic and Highly Dispersive Holograms. *Nano Lett.* **16**, 5235–5240 (2016).

26. M. Khorasaninejad *et al.*, Achromatic Metalens over 60 nm Bandwidth in the Visible and Metalens with Reverse Chromatic Dispersion. *Nano Lett.* **17**, 1819–1824 (2017).

27. E. Arbabi, A. Arbabi, S. M. Kamali, Y. Horie, A. Faraon, Controlling the sign of chromatic dispersion in diffractive optics with dielectric metasurfaces. *Optica*. **4**, 625–632 (2017).

28. E. R. Dowski, W. T. Cathey, Extended depth of field through wave-front coding. *Appl. Opt.* **34**, 1859–1866 (1995).

29. H. B. Wach, E. R. Dowski, W. T. Cathey, Control of chromatic focal shift through wave-front coding. *Appl. Opt.* **37**, 5359–5367 (1998).

30. W. T. Cathey, E. R. Dowski, New paradigm for imaging systems. *Appl. Opt.* **41**, 6080–6092 (2002).

31. A. Zhan, S. Colburn, C. M. Dodson, A. Majumdar, Metasurface Freeform Nanophotonics. *Sci. Rep.* **7**, 1673 (2017).

32. Low-Contrast Dielectric Metasurface Optics - ACS Photonics (ACS Publications), (available at http://pubs.acs.org/doi/abs/10.1021/acsphotonics.5b00660).

33. V. Liu, S. Fan, S4 : A free electromagnetic solver for layered periodic structures. *Comput. Phys. Commun.* **183**, 2233–2244 (2012).

34. J. W. Goodman, *Introduction to Fourier Optics* (Roberts and Company Publishers, 2005).

35. Z. Wang, A. C. Bovik, H. R. Sheikh, E. P. Simoncelli, Image quality assessment: from error visibility to structural similarity. *IEEE Trans. Image Process.* **13**, 600–612 (2004).

36. M. Ohta, K. Sakita, T. Shimano, A. Sakemoto, *Jpn. J. Appl. Phys.*, in press, doi:10.7567/JJAP.54.09ME03.

37. A. W. Lohmann, Scaling laws for lens systems. *Appl. Opt.* **28**, 4996–4998 (1989).

38. Planar metasurface retroreflector : Nature Photonics : Nature Research, (available at http://www.nature.com/nphoton/journal/vaop/ncurrent/full/nphoton.2017.96.html).





39.  F. Heide *et al.*, *ACM Trans Graph*, in press, doi:10.1145/2661229.2661260.

40.  P. Getreuer, Total Variation Deconvolution using Split Bregman. Image Process. Line. 2, 158–174 (2012).


**Acknowledgments:**


**Funding:** This work was facilitated though the use of advanced computational, storage, and networking infrastructure provided by the Hyak supercomputer system at the University of Washington (UW). Part of this work was conducted at the Washington Nanofabrication Facility / Molecular Analysis Facility, a National Nanotechnology Coordinated Infrastructure (NNCI) site at the University of Washington, which is supported in part by funds from the Molecular Engineering & Sciences Institute, the Clean Energy Institute, the Washington Research Foundation, the M. J. Murdock Charitable Trust, the National Science Foundation and the National Institutes of Health. The research work is supported by the startup fund provided by the UW, Seattle, the Intel Early Career Faculty Award, and an Amazon Catalyst Award.

**Author Contributions:** S.C., A.Z., and A.M. conceived the idea. S.C. and A.Z. designed and simulated the devices. S.C. fabricated and characterized the devices and conducted the imaging experiments. S.C., A.Z, and A.M. prepared the manuscript. A.M. supervised the project.

**Competing Interests:** The authors declare that they have no competing interests.